\documentstyle{article}
\begin{document}
\newtheorem{theorem}{Theorem}[section]
\newtheorem{corollary}[theorem]{Corollary}
\newtheorem{lemma}[theorem]{Lemma}
\newtheorem{proposition}[theorem]{Proposition}
\newtheorem{step}[theorem]{Step}
\newtheorem{rmk}[theorem]{Remark}
\newtheorem{que}[theorem]{Question}
\newtheorem{defin}[theorem]{Definition}

\font\sixbb=msbm6
\font\eightbb=msbm8
\font\twelvebb=msbm10 scaled 1095
\newfam\bbfam
\textfont\bbfam=\twelvebb \scriptfont\bbfam=\eightbb
                           \scriptscriptfont\bbfam=\sixbb

\def\bb{\fam\bbfam\twelvebb}
\newcommand{\Rea}{{\bb R}}
\newcommand{\Com}{{\bb C}}
\newcommand{\Int}{{\bb Z}}
\newcommand{\Nat}{{\bb N}}
\newcommand{\FF}{{\bb F}}
\newcommand{\EE}{{\bb E}}
\newcommand{\PP}{{\bb P}}
\newcommand{\rk}{{\rm rank}}
\newcommand{\enp}{\begin{flushright} $\Box$ \end{flushright}}
\def\cD{{\mathcal{D}}}

\title{Wigner symmetries and Gleason's theorem
\thanks{
This research was
supported by grants P1-0288 and J1-2454 from ARRS, Slovenia.}
\author{
Peter \v Semrl\\
Faculty of Mathematics and Physics,
        University of Ljubljana\\
        Jadranska 19,
        SI-1000 Ljubljana,
        Slovenia\\
peter.semrl@fmf.uni-lj.si}
}

\date{}
\maketitle

\begin{abstract}
 We prove an optimal version of Uhlhorn's generalization of Wigner's unitary-antiunitary theorem. The main tool in our proof is Gleason's theorem. 
\end{abstract}
\maketitle



\section{Introduction and statement of the main results}

Let $H$ ba a finite or infinite-dimensional Hilbert space. Throughout the paper we will assume that $H$ is separable and $\dim H \ge 3$. We will denote by ${\cal P}(H)$ the set of all rank one projections on $H$.   For $P,Q \in {\cal P}(H)$ we write $P \perp Q$ if $PQ=0$ which is equivalent to ${\rm tr} (PQ)=0$. For every vector $x \in H$ of norm one we will denote by $P_x$ the rank one projection onto the linear span of $x$. We will say that a subset $\{ P_1 , P_2 , \ldots \} \subset {\cal P}(H)$ is  a complete orthogonal system of projections of rank one, COSP,
if $P_i \perp P_j$ whenever $i \not=j$ and there is no rank one projection $Q$ that is orthogonal to each $P_j$. If $P_j = P_{x_j}$ for some unit vectors $x_j$, $j=1,2, \ldots$, then  $\{ P_1 , P_2 , \ldots \} \subset {\cal P}(H)$ is COSP if and only if $\{ x_1 , x_2, \ldots \}$ is an orthonormal basis of $H$.

Wigner's theorem \cite[pp. 251-254]{Wig}, one of the cornerstones of mathematical foundations of quantum mechanics, states that ever symmetry of the space of pure states is induced by a unitary or antiunitary operator. In the mathematical foundations of quantum mechanics pure states correspond to rank one projections (one may identify rank one projections with one-dimensional subspaces of a Hilbert space) and the transition probability between two pure states corresponds to the trace of the product of the corresponding rank one projections. Thus, in mathematical language Wigner's theorem reads as follows: Let $H$ be a Hilbert space and $\phi : {\cal P}(H) \to {\cal P}(H)$ a bijective map such that
for every pair $P,Q \in {\cal P}(H)$ we have ${\rm tr}\, ( \phi (P) \phi (Q) ) = {\rm tr} \, (PQ)$. Then there exists a unitary or antiunitary operator $U : H \to H$ such that $\phi (P) = UPU^\ast$, $P \in {\cal P}(H)$. Every such map will be called a Wigner symmetry.
Because of its fundamental importance in mathematical physics this result has been reproved and improved in many ways. For some recent results of this type we refer to \cite{Geh1}, \cite{Geh2}, \cite{GeM}, \cite{Mo6}, \cite{Mo7}, and \cite{Pan}.

Wigner's theorem together with many variations serves also as one of the main tools in the study of symmetries of various quantum structures. Assume that we have a bijective map $\phi$ on the set of all self-adjoint operators (bounded observables in the language of quantum mechanics) or on the operator interval $[0,I]$ (the effect algebra in the language of quantum mechanics) or some other quantum structure and suppose that such a map preserves a certain operation and/or a property and/or a relation that are relevant in mathematical physics. The problem is to describe the general form of such maps (symmetries) and one of the most frequently used approach in this kind of problems is to first prove that $\phi$ maps projections of rank one to projections of rank one and preserves the transition probability. Then one can apply Wigner's theorem to conclude that $\phi$ has a nice behaviour when restricted to ${\cal P}(H)$. This is usually the crucial step and then the final step, that is, to show that the symmetry has the desired form on the whole quantum structure, can be technically quite involved but is usually the easier part of the proof. An interested reader can find many examples of such an approach to the study of symmetries in \cite[Chapter 2]{Mol} and the references therein.

When using this approach and restricting $\phi$ to ${\cal P}(H)$ we often do not know wheather $\phi$ maps ${\cal P}(H)$ onto itself. And quite often we know that it preserves only the zero transition probability. And sometimes we know it preservers the zero transition probability in one direction only.
Thus, not all the assumptions of Wigner's theorem are satisfied and quite often we need a much stronger version. This motivates the search for optimal versions of Wigner's theorem.

Recall that a map $\phi : {\cal P} (H) \to {\cal P} (H)$ preserves orthogonality in one direction if for every pair $P,Q \in {\cal P} (H)$ we have $PQ = 0 \Rightarrow \phi (P) \phi (Q) = 0$, and it preserves orthogonality in both directions if for every pair $P,Q \in {\cal P} (H)$ we have $PQ = 0 \iff \phi (P) \phi (Q) = 0$. 
In 1963 Uhlhorn \cite{Uhl} significantly improved Wigner's theorem by showing that assuming  only the preservation of the zero transition probability is enough to get the same conclusion. More precisely, if 
$\phi : {\cal P}(H) \to {\cal P}(H)$ is a bijective map which preserves  orthogonality in both directions, then there exists a unitary or antiunitary operator $U : H \to H$ such that $\phi (P) = UPU^\ast$, $P \in {\cal P}(H)$.

Can we further improve this result? If we start with the weakest possible assumption then we consider maps $\phi : {\cal P} (H) \to {\cal P} (H)$ that preserve orthogonality in one direction only (no surjectivity or injectivity is assumed). It has been recently proved \cite{PaV} that in the special case that $H$ is finite-dimensional such maps are Wigner symmetries. Pankov and Vetterlein used a geometric approach based on a certain
version of the fundamental theorem of projective geometry. Another important tool was Gleason's theorem \cite{Gle} which was needed to prove that orthogonality preserving lineations on the projective space are non-degenerate. 

In the infinite-dimensional case stronger assumptions are needed. It is known \cite{Sem1} that there exist injective maps $\phi : {\cal P} (H) \to {\cal P} (H)$ preserving orthogonality in both directions that behave in a wild way. When dealing with symmetries of effect algebras the following weaker form of Uhlhorn's theorem \cite[Proposition 2.6]{Sem} was obtained: If an injective transformation $\phi : {\cal P} (H) \to {\cal P} (H)$ maps every COSP onto some COSP, then $\phi$ is a Wigner symmetry. It was shown in \cite[Remark 1 and Theorem 2]{PaV} that the same conclusion holds in the absence of the injectivity assumption. Clearly, every transformation $\phi : {\cal P} (H) \to {\cal P} (H)$ which maps every COSP onto some COSP preserves orthogonality in one direction (since every pair of orthogonal rank one projections can be extended to COSP).

Our main result states that it is enough to assume the existence of only one COSP that is mapped onto some COSP to characterize Wigner symmetries.

\begin{theorem}\label{glavni}
Let $\phi : {\cal P} (H) \to {\cal P} (H)$ be a map preserving orthogonality in one direction. Assume that there exists  a complete orthogonal system of projections of rank one $\{ P_1 , P_2 , \ldots \} \subset {\cal P}(H)$ such that $\{ \phi (P_1 ), \phi (P_2 ) , \ldots \}$ is  a complete orthogonal system of projections of rank one. Then $\phi$ is a Wigner symmetry.
\end{theorem}

This result is optimal. We do know that maps on ${\cal P}(H)$ preserving orthogonality in one direction can be wild. So, in order to get a desired conclusion that such a preserver is a Wigner symmetry we need to add a certain regularity condition. The one that we have used seem to be quite natural and obviously cannot be replaced by any weaker assumption.

If $\dim H = n$, $ 3 \le n < \infty$, and $\phi : {\cal P} (H) \to {\cal P} (H)$ preserves orthogonality in one direction then it obviously maps every COSP onto some COSP. Hence, the optimal version of Uhlhorn's theorem in the finite-dimensional case \cite[Theorem 1]{PaV} is a direct consequence of our main result.

\begin{corollary}\label{micj}
\cite{PaV}
Let $3 \le dim H < \infty$. Then every map $\phi : {\cal P} (H) \to {\cal P} (H)$ which preserves orthogonality in one direction is a Wigner symmetry.
\end{corollary}

We will first show that a weaker version of Theorem \ref{glavni} follows rather quickly from Gleason's theorem. This version is strong enough to yield Corollary \ref{micj}. Thus, our first contribution  is a short proof of Pankov-Vetterlein's result. Then we will use this result to deduce our main theorem.

\section{Proofs}

We first recall Gleason's theorem. A density operator $D : H \to H$ is a positive linear operator whose trace is equal to $1$. We will need the following formulation of Gleason's theorem: If $\varphi : {\cal P}(H) \to [0,1]$ is a function such that for every complete orthogonal system of projections of rank one
$\{ P_1 , P_2 , \ldots \}$ we have
$$
\sum_j \varphi (P_j) = 1,
$$
then there exists a density operator   $D : H \to H$ such that
$$
\varphi (P) = {\rm tr}\, (DP)
$$
for every $P\in {\cal P}(H)$.

We will need the following remark concerning Corollary \ref{micj}. The conclusion is that $\phi$ is induced by a unitary or antiunitary operator $U$. Such an operator $U$ is not uniquely determined. But if a Wigner symmetry $\phi$ is induced by $U$ and is also induced by $V$, then either both $U$ and $V$ are unitary, or both $U$ and $V$ are antiunitary. In both cases we have $U = zV$ for some complex number of modulus one. This is well-known and also easy to verify.

We will start by proving the following claim which is a (much) weaker version of Theorem \ref{glavni}: Let $\phi : {\cal P} (H) \to {\cal P} (H)$ be a map such that for every complete orthogonal system of projections of rank one $\{ P_1 , P_2 , \ldots \} \subset {\cal P}(H)$  the set $\{ \phi (P_1 ), \phi (P_2 ) , \ldots \}$ is  a complete orthogonal system of projections of rank one. Then $\phi$ is a Wigner symmetry. 

This result has not been formulated before but it can be deduced easily from \cite[Theorem 2]{PaV}. We will present here a much shorter and simpler proof based on Gleason's theorem and  a non-bijective version of Wigner's theorem. Note that the proof based on the arguments from \cite{PaV} also depends on Gleason's theorem but Wigner's theorem is not involved.

All we need to do to prove the above claim is to
take any density operator $D : H \to H$ and consider the map $\varphi_D : {\cal P}(H) \to [0,1]$ defined by
$$
\varphi_D (P) = {\rm tr}\, (D\phi (P)), \ \ \ P \in {\cal P}(H).
$$
Our assumption yields that Gleason's theorem can be applied. It follows that for every density operator $D: H \to H$ there exists a density operator $E : H \to H$ such that
$$
{\rm tr}\, (D\phi (P)) = {\rm tr}\, (EP), \ \ \ P \in {\cal P}(H).
$$
In particular, if we choose $D = \phi (Q)$ for some fixed $Q \in {\cal P}(H)$, then we have
$$
{\rm tr}\, (\phi (Q)\phi (P)) = {\rm tr}\, (E_Q P), \ \ \ P \in {\cal P}(H),
$$
for some density operator $E_Q$. Choosing $P=Q$ we conclude that
$$
{\rm tr}\, (E_Q Q) = 1.
$$
It is an elementary excercise to see that if $x$ is a unit vector in the image of a rank one projection $Q$ then ${\rm tr}\, (E_Q Q) = \langle E_Q x , x \rangle$. Since $0 \le E_Q \le I$, the standard operator theory arguments show that $E_Q x = x$. Thus, $1$ is an eigenvalue of $E_Q$ and the corresponding eigenspace contains $x$. With respect to the orthogonal direct sum decomposition $H = {\rm span}\, \{ x \} \oplus x^\perp$ the operator $E_Q$ has the matrix representation
$$
E_Q = \left[ \begin{matrix} {1 & 0 \cr 0 & T \cr} \end{matrix} \right]
$$
for some positive operator $T$ acting on the orthogonal complement of $x$. Its trace is zero, and therefore, $T=0$, or equivalently, $E_Q = Q$.

Hence, for every $P \in {\cal P}(H)$ we have
$$
{\rm tr}\, (\phi (Q)\phi (P)) = {\rm tr}\, (Q P).
$$
But $Q$ was chosen arbitrarily. We conclude that $\phi$ preserves the trace of the product (the transition probability). By the non-bijective version of Gleason's theorem (for a simple proof we refer to \cite{Ge0}) there exists a linear or conjugate-linear isometry $U : H \to H$ such that $\phi (P) = UPU^\ast$ for every $P \in {\cal P}(H)$. Using our assumption once more we see that $U$ is surjective, that is, $U$ is either a unitary or antiunitary operator. This completes the proof of the claim.
 
It is clear that Corollary \ref{micj} follows directly from the above claim.

It remains to prove Theorem \ref{glavni}. Because of Corollary \ref{micj} there is no loss of generality in assuming that $H$ is infinite-dimensional. So, let $\phi : {\cal P} (H) \to {\cal P} (H)$ be a map preserving orthogonality in one direction and $\{ P_j \, : \, j\in \Nat \} \subset {\cal P}(H)$ a complete orthogonal system of projections of rank one  such that $\{ \phi (P_j ) \, : \, j\in \Nat \}$ is  a complete orthogonal system of projections of rank one.  After replacing the map $\phi$ by a map $P \mapsto V \phi (P) V^\ast$ for a suitable unitary operator $V$ we may assume with no loss of generality that 
$$
\phi (P_j ) = P_j ,\ \ \ j \in \Nat.
$$
Let $e_j$, $j \in \Nat$, be the corresponding orthonormal basis of $H$, that is, ${\rm Im}\, P_j = {\rm span}\, \{ e_j \}$, $j \in \Nat$.

For any finite subset ${\cal S} \subset \Nat$ with ${\rm card} \, {\cal S} = n \ge 3$ we denote by ${\cal P}_{\cal S}$ the set of all rank one projections $P \in {\cal P} (H)$ satisfying
$$
{\rm Im}\, P \subset \bigoplus_{j \in {\cal S}} {\rm Im}\, P_j  = {\rm span}\, \{ e_j \, : \, j \in {\cal S} \}.
$$
Clearly, ${\cal P}_{\cal S}$ can be identified with ${\cal P} ({\Com^n})$. 

Since $\phi (P_k ) = P_k$ for every positive integer $k$, $k\not\in {\cal S}$, and because $\phi$ preserves orthogonality, we have $\phi ( {\cal P}_{\cal S} ) \subset {\cal P}_{\cal S}$. By Corollary \ref{micj}, there exists a unitary or antiunitary operator $W_{\cal S} : \bigoplus_{j \in {\cal S}} {\rm Im}\, P_j \to \bigoplus_{j \in {\cal S}} {\rm Im}\, P_j$ such that $\phi$ restricted to ${\cal P}_{\cal S}$ is induced by $W_{\cal S}$. It is also clear that there exist complex numbers $z_j$, $j \in {\cal S}$, such that $W_{\cal S} e_j = z_j e_j$, $j \in {\cal S}$. Of course, all $z_j$'s are of modulus one.

From now on we will consider only finite subsets ${\cal S} \subset \Nat$ satisfying $1,2,3 \in {\cal S}$. For every such a subset ${\cal S}$ we will choose a uniquely determined unitary or antiunitary operator  $W_{\cal S} : \bigoplus_{j \in {\cal S}} {\rm Im}\, P_j \to \bigoplus_{j \in {\cal S}} {\rm Im}\, P_j$ which induces the restricion of $\phi$ to ${\cal P}_{\cal S}$ and satisfies $W_{\cal S} e_1 =  e_1$.

Assume that we have two finite subsets ${\cal S}_1 , {\cal S}_2 \subset \Nat$ such that $\{ 1,2,3 \} \subset {\cal S}_1 \subset {\cal S}_2 \subset \Nat$. Then $W_{{\cal S}_1}$ and $W_{{\cal S}_2}$ are either both unitary, or both antiunitary, and 
$W_{{\cal S}_2} e_j = W_{{\cal S}_1} e_j$ for every $j \in {\cal S}_1$. This follows from the fact that a unitary or antiunitary operator inducing the restriction of $\phi$ to ${\cal P}_{{\cal S}_1}$ and mapping $e_1$ to itself is uniquely determined. 

Next, assume that we have two finite subsets ${\cal S}_1 , {\cal S}_2 \subset \Nat$ both of them containing $\{ 1,2,3 \}$. We take ${\cal S} =  {\cal S}_1 \cup {\cal S}_2$ and apply the previous paragraph to conclude that either both $W_{{\cal S}_1}$ and $W_{{\cal S}_2}$ are unitary, or 
both $W_{{\cal S}_1}$ and $W_{{\cal S}_2}$ are antiunitary, and $W_{{\cal S}_2} e_j = W_{{\cal S}_1} e_j$ for every $j \in {\cal S}_1 \cap {\cal S}_2$.

It follows that either all the $W_{\cal S}$'s are unitary, or all the $W_{\cal S}$'s are antiunitary. We will consider just one of the two possibilities, say that all of them are unitary. Then let $W$ be the unique unitary operator such that for every positive integer $j$ we have
$$
W e_j = W_{\cal S} e_j
$$
for some finite subset ${\cal S} \subset \Nat$ such that $\{ 1,2,3, j \} \subset {\cal S}$. If we replace the map $\phi$ by the map $P \mapsto W^\ast \phi (P) W$, $P \in {\cal P}(H)$, then we may assume with no loss of generality that
$$
\phi (P) = P
$$
for every $P \in {\cal P}(H)$ satisfying the property that ${\rm Im}\, P \subset {\rm span}\, \{ e_1 , \ldots, e_q \}$ for some positive integer $q$. We need to verify that $\phi (Q) = Q$ for every $Q \in {\cal P}(H)$.

So, let $Q\in {\cal P}(H)$ be any projection of rank one. Let $x$ be a unit vector that belongs to its image. We choose a unit vector $y$ that belongs to the image of $\phi (Q)$. We want to show that $y=zx$ for some complex number $z$ of modulus one. 

For every unit vector $u$ such that
\begin{itemize}
\item $u \in  {\rm span}\, \{ e_1 , \ldots, e_r \}$ for some positive integer $r$, and
\item $u \perp x$,
\end{itemize}
we have $Q \perp P_u$ which yields $\phi(Q) \perp \phi (P_u) = P_u$, or equivalently, $u \perp y$. The easy verification that this imples  $y=zx$ for some complex number $z$ of modulus one is left to the reader. This completes the proof.

\end{document}